\documentclass[conference]{IEEEtran}
\IEEEoverridecommandlockouts
\usepackage{cite}
\usepackage{amsmath,amssymb,amsfonts}
\usepackage{algorithmic}
\usepackage{graphicx}
\usepackage{textcomp}
\usepackage{xcolor}
\usepackage{verbatim}
\usepackage{diagbox}
\usepackage{multirow}
\usepackage{pbox}
\usepackage{makecell}

\usepackage[english]{babel}

\def\BibTeX{{\rm B\kern-.05em{\sc i\kern-.025em b}\kern-.08em
    T\kern-.1667em\lower.7ex\hbox{E}\kern-.125emX}}
\begin{document}

\makeatletter
\newcommand{\newlineauthors}{%
  \end{@IEEEauthorhalign}\hfill\mbox{}\par
  \mbox{}\hfill\begin{@IEEEauthorhalign}
}
\makeatother

\title{S3M: Siamese Stack (Trace) Similarity Measure}

\author{\IEEEauthorblockN{Aleksandr Khvorov}
\IEEEauthorblockA{\textit{JetBrains, ITMO University} \\
Saint-Petersburg, Russia \\
aleksandr.khvorov@jetbrains.com}
\and
\IEEEauthorblockN{Roman Vasiliev}
\IEEEauthorblockA{
\textit{JetBrains} \\
Saint-Petersburg, Russia \\
roman.vasiliev@jetbrains.com}
\and
\IEEEauthorblockN{George Chernishev}
\IEEEauthorblockA{
\textit{Saint-Petersburg State University} \\
Saint-Petersburg, Russia \\
g.chernyshev@spbu.ru}
\newlineauthors
\IEEEauthorblockN{Irving Muller Rodrigues}
\IEEEauthorblockA{\textit{Polytechnique Montreal} \\
Montreal, Canada\\
irving.muller-rodrigues@polymtl.ca}
\and
\IEEEauthorblockN{Dmitrij Koznov}
\IEEEauthorblockA{\textit{Saint-Petersburg State University} \\
Saint-Petersburg, Russia \\
d.koznov@spbu.ru}
\and
\IEEEauthorblockN{Nikita Povarov}
\IEEEauthorblockA{\textit{JetBrains} \\
Saint-Petersburg, Russia \\
nikita.povarov@jetbrains.com}
}

\graphicspath{ {./figures/} }

\maketitle

\begin{abstract}
Automatic crash reporting systems have become a de-facto standard in software development. These systems monitor target software, and if a crash occurs they send details to a backend application. Later on, these reports are aggregated and used in the development process to 1) understand whether it is a new or an existing issue, 2) assign these bugs to appropriate developers, and 3) gain a general overview of the application's bug landscape. The efficiency of report aggregation and subsequent operations heavily depends on the quality of the report similarity metric. However, a distinctive feature of this kind of report is that no textual input from the user (i.e., bug description) is available: it contains only stack trace information.

In this paper, we present S3M (``extreme'')~--- the first approach to computing stack trace similarity based on deep learning. It is based on a siamese architecture that uses a biLSTM encoder and a fully-connected classifier to compute similarity. Our experiments demonstrate the superiority of our approach over the state-of-the-art on both open-sourced data and a private JetBrains dataset. Additionally, we review the impact of stack trace trimming on the quality of the results.
\end{abstract}


\section{Introduction}


Collection of bug reports is an essential part of the software development process. Bug trackers allow programmers to efficiently concentrate their efforts by prioritizing urgent bugs, discern between new and old bugs, select an appropriate developer for a specific bug, and so on. Overall, bug trackers are a crucial software development tool that greatly improves the efficiency of bug fixing.

Usually, a bug report contains a textual description of the bug, and some categorical metadata such as product version, OS, severity, status etc. Classic bug reports are filled in manually, either by users or product testers. A different approach to bug reporting is an automatic crash reporting system that monitors the target software. If a crash occurs, this system forms and sends a report to a backend application. Unlike classic bug reporting tools, they require no input from a user, containing only a stack trace. 

Such an approach allows to drastically increase the amount of bug feedback at hand, but it is more prone to producing duplicates, thus making deduplication a problem of high priority. Since such reports lack textual content, conventional deduplication techniques (for example, see surveys in studies~\cite{8919115,Hindle2018,8048025,6100061}) are inapplicable. Thus, the stack trace-based report deduplication problem gives rise to a separate class of approaches that address it with a similarity measure; and having a good similarity measure has a massive impact on the quality of report bucketing. This directly affects the quality of decision-making in the software development process.

This problem has been intensively studied for more than a decade, spawning a significant number of papers. Existing approaches rely on two crucial ideas: string similarity (e.g. stack frame-based Levenshtein distance)~\cite{brodie, bartz, dhaliwal, modani} and information retrieval approaches (e.g. TF-IDF applied to stack frames)~\cite{Lerch:2013:FDY:2495256.2495763, Campbell:2016:UET:2901739.2901766, Wu:2014:CLC:2610384.2610386}. 

Recently, Deep Learning (DL) techniques have been successfully adopted for the classic report deduplication problem~\cite{8094414, 10.1145/3297001.3297023, 10.1145/3379597.3387470, DWEN}. However, to the best of our knowledge, there are still no approaches that apply DL to stack trace-based report deduplication. At the same time, DL techniques look promising for three reasons:
\begin{enumerate}
    \item While processing a stack trace, DL techniques can take into account frame context (other frames of the stack trace). This means that neural networks can, for example, ``assess'' the fifth frame while ``thinking'' of the first one. Existing (classic) methods for stack trace-based report deduplication are not capable of this as they process stack frames independently. 

    \item Embeddings can allow neural networks to infer semantic similarity of methods in stack traces. For example, a model can ``understand'' that methods \textit{get()} and \textit{getAll()} are related to each other.
    \item Neural networks do not require manual feature engineering; instead, they are able to perform feature extraction in an automatic and trainable way. Therefore, classic approaches are limited to features that were built-in by their designers, while DL approaches can potentially learn more complex ones, thus resulting in increased performance and reduced dataset dependence.
    
\end{enumerate}

\begin{figure*}[!ht]
\centering
\includegraphics[width=0.8\linewidth]{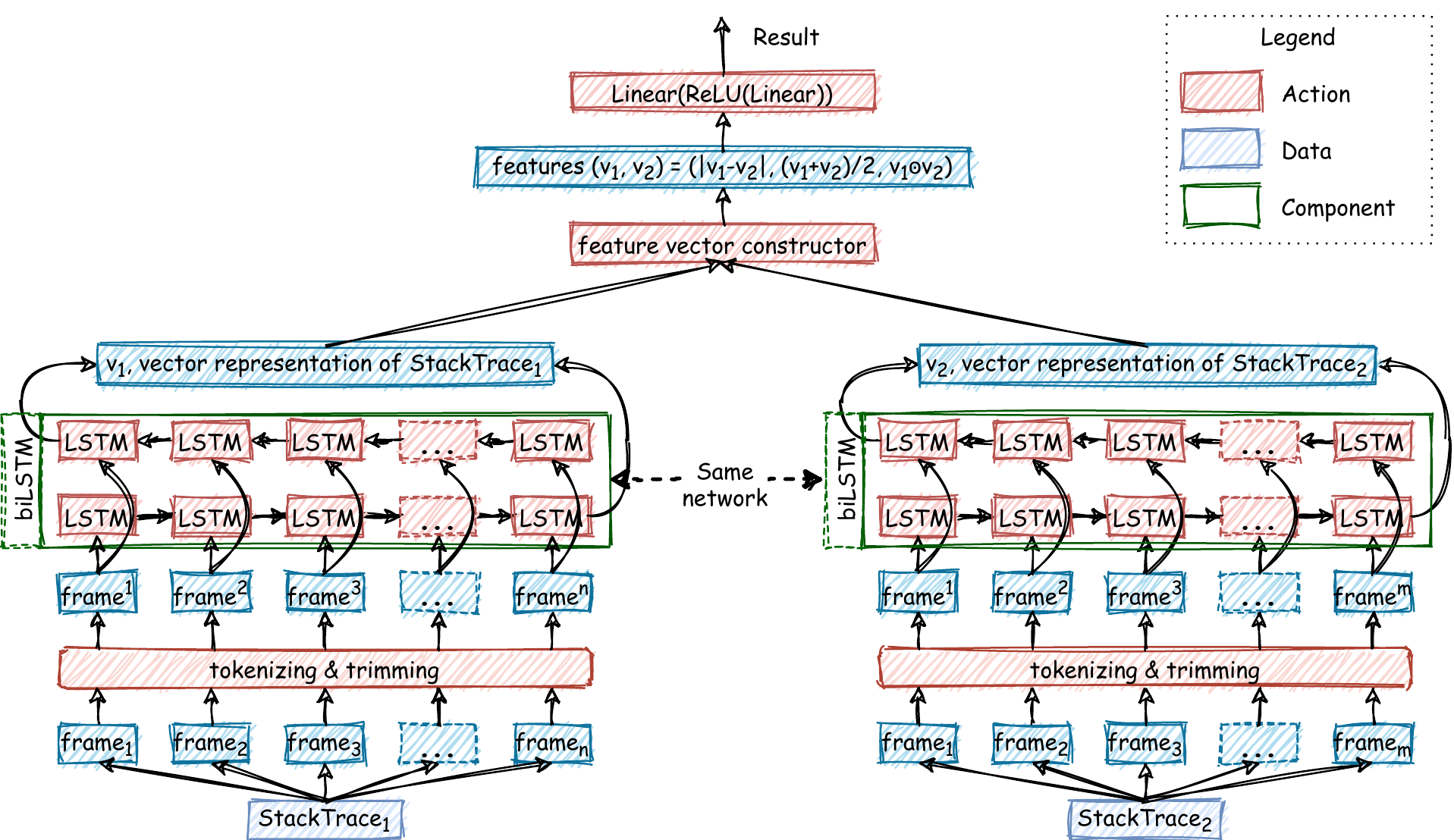}
\caption{Approach Architecture}
\label{fig:algo}
\end{figure*}


In this paper we present our model~--- S3M (pronounced ``extreme'')~--- the first DL approach to calculating stack trace similarity for crash report deduplication. It is based on a siamese architecture that uses a biLSTM encoder and two fully-connected layers with ReLU activation as classifier.

We experimentally evaluate our model on an open-source dataset and private JetBrains data. Preliminary experiments demonstrate the superiority of our approach over classic (non-DL) approaches to stack trace-based report deduplication. Additionally, we demonstrate the beneficial impact of stack frame trimming~--- a technique that we had originally proposed to reduce the number of tokens in the dictionary. 


\section{Related Work}\label{sec:relwork}

Stack trace-based report deduplication is a well-established area. The first studies from the mid-00's approached the problem via string matching algorithms: Brodie et al.~\cite{brodie} adapted a biological sequence searching algorithm; Modani et al.~\cite{modani} compared edit distance, prefix match and LCS approaches; Bartz et al.~\cite{bartz} presented a failure similarity classifier that takes custom edit distance between callstacks as one of the parameters; Dhaliwal et al.~\cite{dhaliwal} proposed a two-level grouping scheme with Levenshtein distance on the second level; Dang et al.~\cite{dang} described the ReBucket technique, in which they have employed a stack trace similarity measure based on string matching for agglomerative hierarchical clustering.

Later, approaches that employ information retrieval techniques have appeared: Lerch and Mezini~\cite{Lerch:2013:FDY:2495256.2495763} 
proposed to adapt TF-IDF for stack frames; and Campbell et al.~\cite{Campbell:2016:UET:2901739.2901766} ran a comparison of various approaches to implementing TF-IDF methods; Sabor et al.~\cite{DURFEX} presented DURFEX, which idea is to consider stack frames as N-grams of variable length after trimming them to package level.

Finally, both Moroo et al.~\cite{DBLP:conf/seke/MorooAH17} and Vasiliev et al.~\cite{10.1145/3416505.3423561} describe the most recent methods, which are both based on the idea of combining TF-IDF and string matching approaches.

\section{Approach}

Our approach relies on LSTM (Long Short-Term Memory)~--- a recurrent neural network. The idea is to represent stack traces as sequences that contain tokens from a predefined dictionary. This approach is similar to ones used for texts and allows to adopt many models that were developed in natural language processing domain. We employ a siamese architecture\cite{10.1109/CVPR.2005.202}~--- a model in which both stack traces are encoded by a single network.

The overall scheme of the proposed algorithm is presented in Figure~\ref{fig:algo}. First, we describe the data preprocessing and stack frame trimming. Then, we describe the siamese LSTM that is used to obtain the representation for a pair of stack traces. Finally, the trained features are used to compute the resulting similarity.


\begin{table*}
\centering
\caption{Dataset details}
\begin{tabular}{l|l|l|l|l|l|l} 
\cline{1-7}
\multicolumn{1}{l|}{\multirow{2}{*}{}} & \multicolumn{3}{c|}{JetBrains} & \multicolumn{3}{c}{NetBeans}  \\ 
\cline{2-7}
\multicolumn{1}{l|}{}                  & Train & Validation & Test            & Train & Validation & Test      \\ 
\hline
Buckets                              &   8769    &  273   &   916          & 31349    &   1592   &    5909           \\ 
\hline
Reports                                &   299444    &   7567   &    56631             & 39789    &   1976   &    7792          \\
\hline
Date                                &  08/09/2018     &  11/22/2018    &  11/29/2018               &  09/25/1998     &   03/26/2010         &   08/13/2010        \\
\hline
\end{tabular}
\label{tbl:dataset}
\end{table*}

\subsection{Preprocessing: Tokenization and Trimming}

We have to transform stack traces into vectorized representations. In our work, a stack trace is represented as a sequence of frames $ST = \{f_n, f_{n-1}, \ldots, f_1\}$. In this sequence, $f_i$ is the $i$-th stack frame.  


Similarly to DURFEX~\cite{DURFEX} we have decided to apply stack frame trimming as the preprocessing step of our approach. The idea is the following: suppose that we have a $com.intellij.psi.impl.source.PsiFileImpl.getStubTree$ frame. By trimming it to, for example, package level, we obtain $com.intellij.psi.impl.source$. This technique allows to reduce the number of tokens that we have to keep in the dictionary. In the current work, we have decided to try different types of trimming (e.g. function, class, package). Aside from reducing the dictionary, this approach should improve the quality of the similarity since the neural network will more frequently encounter an individual frame, thus giving it more chances to learn it. 

\subsection{Vector Representation of Stack Traces}

Our goal is to design a stack trace similarity measure that will use neural networks. This measure should be symmetric: $similarity (stack_1, stack_2) = similarity (stack_2, stack_1)$.

To ensure this property, we have decided to follow the siamese approach and use the appropriate feature vector (see below). In the siamese approach, a shared neural network independently encodes each input (stack traces) into a vector representation. 
The siamese approach has been successfully applied in different domains~\cite{poddar-etal-2019-train, 10.1145/2766462.2767738, Yushi2016} and it can be efficient. Instead of generating stack trace representations for each comparison, it is possible to store them and only execute the classifier. Since representation generation is the most computationally expensive part of S3M, this in turn reduces the generation time of the recommendation lists. Additionally, using a siamese network allows to reduce the number of parameters to be set during training.

To handle sequential data, an RNN architecture called LSTM~\cite{10.1162/neco.1997.9.8.1735} is frequently used. These networks sequentially iterate over data items (tokens) and on each step they update their internal state and produce the resulting vector. Essentially, this allows the RNN to ``remember'' previous context and consider it while processing the next token. However, this approach has an obvious drawback: it takes into account only tokens that were shown to the RNN before the considered term. At the same time, subsequent terms may define the correct output of the RNN.

To address this issue, we have decided to use the biLSTM~\cite{10.1007/11550907-126} approach. It is, essentially, two LSTMs, one of which accepts a token sequence in a direct order, and the other in a reversed one. The resulting vector is obtained by concatenating their final outputs (i.e. ones that are produced after processing the final token).

Next, we construct the feature vector that will be used by our neural network classifier. Similarly to study~\cite{Yushi2016}, we have decided to construct is as follows:

\vspace{0.10cm}
    ${features(v_1, v_2)} = \left(\left|v_1 - v_2\right|, \left(v_1 + v_2\right) / 2, \;v_1 \odot v_2 \right)$,
\vspace{0.10cm}

where $v_1 = biLSTM(stack_1)$, $v_2 = biLSTM(stack_2)$. In this formula, component-wise subtraction, arithmetic mean computation, and multiplication is used. The thought process behind such an approach is that features should represent differences between original vectors as well as similarities.

\subsection{Algorithm}


Having obtained encoded representations of both stacks, we feed them into another neural network which will predict the resulting similarity:

\vspace{0.10cm}
$similarity = Linear(ReLU(Linear(features)))$.
\vspace{0.10cm}

Essentially, it is a fully connected 2-layer network with ReLU activation which returns the similarity of two stack traces. Due to space constraints, we do not give detailed descriptions of the DL concepts used in this section. However, they can be found in~\cite{10.5555/3086952}.


\subsection{Training}

We have used the RankNet loss for training in order to better take into account the ranking nature of our problem. As a relevant answer, we took a random stack trace from a correct bucket. We generated 4 non-relevant answers from a random subset made up from stack traces that belong to the top 50 buckets selected by Lerch and Mezini~\cite{lerch} (except the correct one). Such an approach to negative sampling can train the model to better discern between similar stack traces rather than just selecting a random stack trace pair. Neural network and embeddings were randomly initialized and trained in an end-to-end manner.

We have used the Adam optimizer~\cite{DBLP:journals/corr/KingmaB14} with a learning rate of $1e-4$. Dimensions for embeddings were set as $50$, and the LSTM hidden size was $100$. The hyperparameters were selected on the validation set for trim = 0 and fixed for other trim levels.


\begin{table*}
\centering
\caption{Results: Comparison with state-of-the-art (top part) \& Effects of trimming (bottom part)}
\begin{tabular}{l|l|l|l|l|l|l|l|l} 
\hline
\multirow{2}{*}{\diagbox{Method}{Metric}} & \multicolumn{4}{c|}{JetBrains} & \multicolumn{4}{c}{NetBeans}   \\ 
\cline{2-9}
                                                   & MRR & RR@1 & RR@5 & RR@10      & MRR & RR@1 & RR@5 & RR@10        \\ 
\hline
Prefix Match~\cite{modani}                         &  0,70   &   0,60   &   0,81   &    0,86        &  0,35   &   0,30   &   0,40   &    0,42        \\ 
\hline
Brodie et al.~\cite{brodie}                        &  0,70   &   0,61   &   0,81   &    0,84        &  0,46   &   0,38   &   0,55   &   0,59           \\ 
\hline
Rebucket~\cite{dang}                               &  0,72   &   0,62   &   0,84   &     0,86       &  0,45   &   0,38   &   0,54   &    0,57         \\ 
\hline
Lerch and Mezini~\cite{Lerch:2013:FDY:2495256.2495763}                                              &  0,76   &   0,65   &   0,90   &      0,92      &  0,51   &  0,41    &   0,64   &     0,69  \\ 
\hline
Moroo et al.~\cite{DBLP:conf/seke/MorooAH17}       &  0,78   &   0,68   &   0,89   &      0,92      &  0,47   &   0,39   &   0,56   &    0,60      \\ 
\hline
DURFEX~\cite{DURFEX}                                             &  0,79   &   0,70   &   0,92   &    0,94        &  0,56   &   0,44   &   \textbf{0,72}   &   \textbf{0,77}  \\ 
\hline
TraceSim~\cite{10.1145/3416505.3423561}            &  0,81   &   0,72   &  0,92    &      0,94      &  0,50   &   0,42   &   0,59   &    0,62      \\ 
\Xhline{3\arrayrulewidth}
S3M (our method)                                           &  \textbf{0,86}   &   \textbf{0,77}   &   \textbf{0,96}   &    \textbf{0,96}        &  \textbf{0,62}   &  \textbf{0,53}    &   \textbf{0,72}   &   0,76                   \\ 
\hline
\hline
\,\,\,\,\,trim = 0 (function)                                             &   0,86  &   0,77   &   0,96   &   0,96         &   0,53  &   0,45   &   0,62   &   0,66               \\ 
\hline
\,\,\,\,\,trim = 1 (class)                                             &  0,85   &   0,76   &   0,95   &   0,96       &  0,62   &  0,53    &   0,72   &  0,76    \\ 
\hline
\,\,\,\,\,trim = 2 (package)                                             &  0,84   &   0,74   &  0,95    &     0,96       &  0,57   &   0,45   &   0,71   &     \textbf{0,77}             \\

\hline
\,\,\,\,\,trim = 3                                             &  0,83   &   0,74   &  0,95    &     0,96       &  0,51   &   0,40   &   0,66   &     0,72             \\


\end{tabular}
\label{tbl:exp}
\end{table*}

\section{Experiments}


We have compared our approach with state-of-the-art methods selected from the ones mentioned in the Related Work section. We posed the following research questions (RQs):

\textbf{RQ1:} How well does our approach perform compared to the state-of-the-art approaches?

\textbf{RQ2:} How is the performance of our approach impacted by different depth (level) of stack frame trimming?

Following the motivation presented in recent report triaging studies (e.g. ~\cite{10.1145/3379597.3387470, DURFEX, lerch}), we have adopted two ranked retrieval metrics: Recall Rate (RR@k) and the MRR metric.

Experiments were run on two datasets: a private JetBrains dataset, and an open-sourced one (NetBeans). The JetBrains dataset was created from stack traces that emerge after every crash of a JetBrains product. The NetBeans dataset was generated from reports submitted before 2016 in the bug tracker of NetBeans's project\footnote{https://bz.apache.org/netbeans}. Crash reports were created by extracting stack traces from the description field and attached files in the bug reports using the regular expression proposed by Lerch and Mezini~\cite{lerch}. To ensure the reproducibility of our results, we have released the NetBeans dataset\footnote{https://figshare.com/articles/dataset/netbeans\_stacktraces\_json/14135003} and the implementation\footnote{https://github.com/akhvorov/S3M} of our approach.

Our experiment methodology is as follows: we have selected a sequential interval of reports and performed a time-aware split~\cite{8919115} into three parts: train, validation, and test. The train and test are classic, and validation is a special fragment of the dataset that is used for tuning the hyperparameters of algorithms. For methods that do not need hyperparameter tuning, the validation step is skipped. 
Train, validation, and test duration were 105, 7 and 35 days for the JetBrains dataset, and 4200, 140, and 700 days for NetBeans. We have selected the dates as to be a multiple of a week, since it is the basis of the work schedule which may impact report arrival rates. We have selected a smaller time frame for the JetBrains dataset since their report arrival rate is much higher. Dataset details are presented in Table~\ref{tbl:dataset}.

For \textbf{RQ1}, the results are presented in the top part of Table~\ref{tbl:exp}. Our approach demonstrates superior results the JetBrains dataset for both MRR and RR@k. For the NetBeans dataset, the outcome is a bit different: our approach wins on MRR and RR@1-5 by a significant margin, but loses considerably on RR@10. Therefore, our method is more preferable for building a fully automated deduplication system. At the same time, DURFEX is more appropriate for building a recommender system.

Finally, we can see that the majority of well-performing approaches rely on TF-IDF, and string matching approaches are considerably worse. 

There is a noticeable difference in the results for all approaches including S3M on JetBrains and NetBeans datasets. This happens due to the nature of the data: the JetBrains dataset has much larger buckets~--- $35,5$ stack traces on average, and NetBeans has a lot of smaller ones~--- $1,3$ stack traces on average. Therefore, NetBeans buckets are less successful in ``attracting'' new stack traces (i.e. this dataset is more ``sparse'').



Next, we have addressed \textbf{RQ2}. For this, we have run our approach with different levels of trimming: function, class, package, and the one with depth 4. The results are presented in Table~\ref{tbl:exp}, in the bottom part. It shows that trimming helps the NetBeans dataset, but it does not improve the results on the JetBrains data. At the same time, the NetBeans dataset demonstrates that the optimal trimming value lies somewhere between class and package. We suppose that there is no ``universal'' level of trimming which should be applied to all datasets. In other words, the trimming level is a hyperparameter of our algorithm, and it should be selected on a per-dataset basis.


\section{Threats to Validity}
Our study has the following threats to validity:

\begin{itemize}
	\item \textbf{Subject selection bias.} The performance of machine learning algorithms frequently depends on the data. Thus, applying our algorithm in other projects may yield different results. We tried to mitigate this by validating it on several different datasets: an open-sourced project and JetBrains product data.
	
	\item \textbf{Limited scope of application.} Our approach is intended for mass-deployed applications, which defines its scope of application: 1) we assume a substantial stream of reports, and 2) we assume that only stack trace information is available, there is no user input, i.e. textual descriptions. 
	
	\item \textbf{Programming language bias.} Our approach was evaluated for stack traces that are produced by the JVM languages. Exceptions of other languages may yield different results. For example, stack trace trimming may have to be altered for C++ exceptions, since C++ stack frames are generally shorter. We leave extending our study to other languages to future work.

\end{itemize}

\section{Conclusions and Future Work}

In this paper we have presented the first approach to computing stack trace similarity using deep learning techniques. Our approach is based on a siamese architecture that uses biLSTM as an encoder and two fully-connected layers with ReLU activation  to compute stack trace similarity. Our experiments have demonstrated the superiority of our approach over the state-of-the-art on both a private JetBrains dataset and open-source NetBeans data. We have also studied the impact of stack trace trimming on the quality of results.

Our future work will include: 1) a more detailed study of alternative feature representations and architectures, effects of trimming, and stack traces of languages other than Java, and 2) the application of transformer networks like GPT-3 and BERT.
\section*{Acknowledgments}


We would like to thank Anastasia Miller for her invaluable comments. Furthermore, we would like to thank Anna Smirnova for her extensive help in the preparation of the present article.







\bibliographystyle{IEEEtran}
\bibliography{main}

\end{document}